# Realization of the field-free Josephson diode

Heng Wu[1,2,4,6]*, Yaojia Wang[1,4,6], Yuanfeng Xu[1,5], Pranava K. Sivakumar[1], Chris Pasco[3], Ulderico Filippozzi[4], Stuart S. P. Parkin[1], Yu-Jia Zeng[2], Tyrel McQueen[3], Mazhar N. Ali[1,4]*

[1]Max Planck Institute of Microstructure Physics, Halle, Saxony-Anhalt, 06108, Germany

[2]College of Physics and Optoelectronic Engineering, Shenzhen University, Shenzhen 518060, China

[3]Johns Hopkins University, Baltimore, Maryland 21218, USA

[4]Kavli Institute of Nanoscience, Delft University of Technology, Delft, the Netherlands

[5]Department of Physics, Princeton University, Princeton, New Jersey 08544, USA

[6]These authors contributed equally: Heng Wu, Yaojia Wang

*e-mail: wuhenggcc@gmail.com, maz@berkeley.edu

Abstract:

**The superconducting analog to the semiconducting diode, the Josephson diode, has long been sought with multiple avenues to realization proposed by theorists[1-3]. Exhibiting magnetic field free, single directional superconductivity with Josephson coupling, it would serve as the building block for next generation superconducting circuit technology. Here we realized the Josephson diode by fabricating an inversion symmetry breaking van der Waals heterostructure of $NbSe_2/Nb_3Br_8/NbSe_2$. We demonstrate, for the first time without magnetic field, that the junction can be superconducting with positive current while resistive with negative current. The $\Delta I_c$ behavior (the difference between positive and negative critical currents) with magnetic field is symmetric and Josephson coupling is proven through the Fraunhofer pattern. Also, stable half-wave rectification of a square-wave excitation was achieved with a very low switching current density, high rectification ratio and high robustness. This nonreciprocal behavior strongly violates the known Josephson relations and opens the door to discover new mechanisms and physical phenomena through integration of quantum materials with Josephson junctions, and provides new avenues for superconducting quantum devices.**

Reciprocity in charge transport describes the symmetric behavior of current with voltage; the magnitude of the current generated by a positive voltage is the same as that generated by a negative voltage[4]. Violating reciprocity, i.e. nonreciprocal behavior, is the basis for numerous important electronic devices such as diodes, rectifiers, AC/DC converters, photodetectors, transistors and more[5,6]. A well-known example of a nonreciprocal device is the p-n junction, formed by the interface of a p-doped and n-doped semiconductor, which exhibits an asymmetric current-voltage characteristic (IVC) without applied magnetic field ($\boldsymbol{B}$) and is widely used in various semiconducting technologies including logic and computation[7].

So far, in the superconducting regime, field-induced nonreciprocal behavior has been seen in noncentrosymmetric superconductors[8-10] and two dimensional electron gases[11], and multiple mechanisms for this phenomenon, including magnetochiral anisotropy[9], finite momentum superconductivity[12], and nonreciprocal Landau critical momentum[13] have been proposed. All require applied magnetic field to achieve simultaneous breaking of inversion symmetry ($\mathcal{P}$) and time reversal symmetry ($\mathcal{T}$) [14-16] so that with sufficient magnetic field strength, the critical current to destroy the superconducting state in one direction can be different from that in the other direction[8]. Nonreciprocal superconductivity *without* an applied magnetic field as not yet been observed either in a bulk superconductor or a superconducting junction, like a Josephson junction.

In particular, the Josephson junction (JJ) is known to exhibit reciprocal behavior as well, as can be seen by the Josephson relations[17,18]. Josephson junctions are typically formed by two superconducting electrodes bridged by a nonsuperconducting barrier comprised of classical materials like $AlO_x$, Cu, etc[19,20]. However, integrating novel quantum materials as the barriers in JJs may lead to anomalous behavior violating the Josephson relations (like nonreciprocal behavior) and is of fundamental and technological interest. Creating a superconducting analog to the semiconducting diode, a field-free "Josephson diode" (JD), allowing supercurrent in one direction but normal current in the other direction in the absence of applied magnetic field, is an open challenge.

One approach is to look for nonreciprocal IVCs in superconducting systems that only break $\mathcal{P}$ but not $\mathcal{T}$. Recently, Josephson junctions with only $\mathcal{P}$ breaking were predicted to exhibit asymmetric IVCs due to asymmetric charging energy[1] or the formation of a Mott-insulating region at the interface of p-doped and n-doped superconductors[2]. Since JJs are the basis of a variety of important technologies[21,22] including SQUIDs (superconducting quantum interference devices), superconducting quantum computing bit (Qubits), and rapid single flux quantum (RSFQ) devices[23-25], the realization of Josephson diode will have a significant impact on superconducting electronics.

Here we report field-free nonreciprocal transport behavior in a Josephson diode of a $\mathcal{P}$-breaking van der Waals (vdW) heterostructure of $NbSe_2/Nb_3Br_8/NbSe_2$. An asymmetric IVC, with sharp superconducting transitions, is seen at zero magnetic field indicating its field-free JD behavior. A superconducting half-wave rectification of a square-wave excitation is demonstrated for the first time, with a large rectification ratio, ultralow switching current density, and highly stable performance. Moreover, the $\Delta I_c$ vs $\boldsymbol{B}$, presents a symmetric behavior, unlike the antisymmetric response seen in previous systems[8], further proving its field-free nature. The field-free JD effect may lie in the asymmetric tunneling of the supercurrent across the tunnel barrier, and the $\mathcal{P}$-breaking of the junction is confirmed by antisymmetric second harmonic peaks. The effect shown here may be extended to a variety of material systems and architectures, particularly utilizing vdW materials, for the advancement of future superconducting electronics and quantum devices. Also, this study shows how the interplay between symmetry breaking, novel tunnel barrier materials and superconducting electrodes can lead to novel or anomalous Josephson phenomena in JJs, opening the door to a variety of fundamental and applied research directions.

We fabricated vertical JJs by stacking thin-flake NbSe$_2$ on either side of an Nb$_3$Br$_8$ thin flake as shown in Fig. 1a. Nb$_3$Br$_8$ is a vdW quantum material that crystallizes in the $R\overline{3}m$ space group at low temperature and its unit cell is comprised of 6 layers of Nb-Br edge-sharing octahedra[26,27] with inversion centers lying between adjacent layers (Extended Data Fig. 1a). The inversion symmetry is preserved in even-layered Nb$_3$Br$_8$ but broken in odd-layered Nb$_3$Br$_8$. Additionally, Nb$_3$Br$_8$ was shown to have a singlet magnetic ground state[27] and moderate band gap like its sister material, Nb$_3$Cl$_8$[27-30]. (See Supplementary Information (SI) Section 1 for further discussion of the magnetoresistance characterization of the Nb$_3$Br$_8$ thin flakes). The NbSe$_2$/Nb$_3$Br$_8$/NbSe$_2$ heterostructures were fabricated by the dry transfer method[31] (see Methods) and capped by a top h-BN to avoid degradation, as a typical device in the inset of Fig. 1b shows. Here NbSe$_2$ flakes with thickness about 28 - 37 nm are used as superconducting electrodes, rotated by arbitrary angles with respect to each other, breaking inversion symmetry of the JJ, and a thin flake of Nb$_3$Br$_8$ with thickness of 2.3 nm (3 layers) is used as the tunnel barrier.

While cooling the junction to 20 mK in zero field, a sudden drop to zero resistance at $T_c$ = 6.6 K is observed, as shown in Extended Data Fig. 1b. Figure 1b investigates the voltage vs. current ($V$-$I$) behavior, measured by sweeping the DC current at 20 mK (see $V$-$I$ curves at different temperatures in Extended Data Fig. 1c). We emphasize that the measurement of the $V$-$I$ curves contains four branches; sweeping the current from zero to a positive value (0-p), from a positive value back to zero (p-0), from zero to a negative value (0-n) and from a negative value back to zero (n-0). These four branches can be measured in the above order (defined as the positive sweep), or in reverse order, 0-n, n-0, 0-p, and p-0 (defined as the negative sweep). Figure 1b shows $V$-$I$ curves of the positive sweep and the negative sweep; both of them exhibit hysteresis, as expected, arising from the capacitance induced different critical currents when breaking ($I_c$) compared with returning ($I_r$) to the superconducting state[1]. Note that the properties measured here all come from the JJ itself as the maximum current applied here is much smaller than the critical current of thick NbSe$_2$ (few milliamps)[32]. Importantly, the two curves lie precisely on top of each other meaning that the positive versus negative sweep order does not affect the $V$-$I$ response. However, the $I_c$ and $I_r$ in the positive current regime (labelled $I_{c+}$ and $I_{r+}$) can be compared with those in the negative current regime (labelled $I_{c-}$ and $I_{r-}$). For reciprocal transport, as is expected for a conventional JJ without magnetic field, there should be no difference between $I_{c+}$ vs. $|I_{c-}|$ or $I_{r+}$ vs. $|I_{r-}|$.

Figure 1c shows the absolute values of $I_{c+}$ and $I_{r+}$ as well as $I_{c-}$ and $I_{r-}$ of the positive sweep. Immediately evident is that $I_{c+}$ and $I_{r+}$ are *not equal* to their negative current counterparts, with a $\Delta I_c$ ($\Delta I_c = |I_{c-}| - I_{c+}$) ~ 0.5 μA and a smaller $\Delta I_r$ ~ 0.1 μA, indicating the nonreciprocal $V$-$I$ response in our JJ. It is important to note that these differences are intrinsic properties of the JJ rather than an extrinsic

Joule heating effect (see discussion in Methods). The different absolute values of $I_{c+}$ and $I_{c-}$ combined with the sharp superconducting transition imply that when the applied current is in between $I_{c+}$ and $|I_{c-}|$, the junction would exhibit superconducting behavior with a negative current but normal conducting behavior with a positive current.

Based on this idea, we demonstrated half-wave rectification at zero field, as shown in Fig 2a. Due to the $I_{c+}$ and $|I_{c-}|$ being 7.61 µA and 8.14 µA, respectively, we applied a square-wave excitation (Fig 2a top panel) with an amplitude of 7.9 µA at a frequency of 0.1 Hz. As shown in the bottom panel of Fig 2a, the junction remains in the superconducting state with the negative current (purple area) while switching to the normal state during the positive current (white area). The measured voltage in the superconducting state is smaller than $4 \times 10^{-7}$ V (the lowest detectable limit of our instrument), while the normal state junction voltage is 1.6 mV, meaning the diode rectification ratio demonstrated here is at least $\sim 10^4$. In addition, the switching current density and switching power are $2.2 \times 10^2$ A/cm$^2$ and 12.3 nW, respectively, which are 2 and 4 orders of magnitude smaller than in the bulk superconducting diode[8].

We further probed the durability of the JD and robustness of half-wave rectification, as shown in Fig. 2b by conducting 10,000 continuous cycles with an applied square-wave excitation with an amplitude of 7.9 µA at 0.5 Hz and zero field. We note that the "on" and "off" state voltages stayed constant during the 10,000 cycles, and the device remains stable after that, indicating the high stability of the JD. Importantly, the JD effect is not just an ultralow temperature property; according to the asymmetric *V-I* curves observed at 0.9 K and 3.86 K (see Extended Data Fig. 2), an ideal half-wave rectification was also realized at 0.9 K while the half-wave rectification at 3.86 K had some punch-through error likely due to the thermal fluctuation as shown in Extended Data Fig. 2b and 2d, respectively. In addition, the large hysteresis at ultralow temperatures indicates the junction is in the strongly underdamped regime. However, by 3.86 K, the p-0 (n-0) almost coincides with the 0-p (0-n) branch, indicating the junction changed to the overdamped regime (see SI Section 4 for more discussion), allowing possible operation frequencies on the order of the superconducting switching speed (THz)[24,33].

In order to further confirm the *field-free* nature of the JD effect, we measured *V-I* curves containing 0-p and 0-n branches with applying an in-plane magnetic field (inset of Fig. 3a) from 40 mT to -40 mT (sweep down) with a step of 0.5 mT. As before, the $I_{c+}$ and $|I_{c-}|$ was extracted and plotted in Fig. 3a (*V-I* curves with magnetic field were shown in Extended Data Fig. 3). Both $I_{c+}$ and $|I_{c-}|$ decrease with increasing magnetic field, and reach $\sim 0$ by 40 mT, which reflects the magnetic field dependent critical currents of a JJ and will be addressed in Fig. 4. Below 35 mT, $\Delta I_c$ emerges, as the $|I_{c-}|$ increases faster than $I_{c+}$ as $\boldsymbol{B}$ goes to 0. Note that $|I_{c-}|$ stays larger than $I_{c+}$ regardless of the direction of the magnetic field. Fig. 3b shows the $\Delta I_c$ plotted as a function of $\boldsymbol{B}$; the $\Delta I_c$ is sustained at zero field and its magnitude is independent of $\boldsymbol{B}$ in the low-field region. The inset of Fig. 3b shows the variation of $\Delta I_c$ with decreasing $|I_{c-}|$, and that $|I_{c-}|$ can be up to 25% larger than $I_{c+}$. The field dependent $\Delta I_c$ behavior here differs from previously mentioned field-induced superconducting diode effect where the $\Delta I_c$ exhibits an antisymmetric field dependence[8]. Also, in bulk superconductors with field-induced superconducting diode effects, $\Delta I_c$ vs $\boldsymbol{B}$ has been sweep direction dependent, flipping sign[8], while in the field-free JD demonstrated here, $\Delta I_c$ vs $\boldsymbol{B}$ shows no dependence on sweep direction (see Extended Data Fig. 4).

To prove Josephson coupling of the junction, the d$V$/d$I$ mapping as a function of bias current and magnetic field was measured and is shown in Fig. 4a. As expected, the characteristic single slit Fraunhofer pattern from the Josephson effect is observed (see SI Section 4 for additional measurements and discussion). The critical current of the JJ decreases rapidly with increasing magnetic field, consistent with the $I_c$ decreasing in Fig. 3a. To further confirm the $\mathcal{P}$-breaking in the junction, the field dependent first and second harmonic resistances were measured with an AC current with an amplitude of 5 µA and frequency of 7.919 Hz, as shown in Fig. 4b. The sharp drop of the first harmonic resistance in ~±30 mT shows the onset of superconductivity while the two antisymmetric peaks appear in the second harmonic resistance at the same fields indicating that inversion symmetry in the junction must be broken[8-10]. There can be two contributions to $\mathcal{P}$-breaking here; first is the intrinsic lack of inversion symmetry of the 3-layer $Nb_3Br_8$, and second is the geometry of the vertical junction where the top and bottom $NbSe_2$ electrodes can be arbitrarily rotated relative to each other and the $Nb_3Br_8$ layer. Hence we also fabricated and measured 4-layer $Nb_3Br_8$ devices, which show the same symmetric $\Delta I_c$ vs $B$ behavior and field-free JD effect as the 3-layer $Nb_3Br_8$ JJ as well as antisymmetric peaks in the second harmonic resistance measurements (see details in SI Section 2). These results indicate that the asymmetric $NbSe_2/Nb_3Br_8$ interfaces in the junction play an essential role for inversion symmetry breaking and the field-free JD effect.

Finally, we discuss the possible origin of $\Delta I_c$ and the diode behavior. Fundamentally, to obtain a $\Delta I_c$ and $\Delta I_r$, an asymmetric $V$-$I$ curve is required i.e. the $I_+$ behavior needs to be distinct from the $I_-$ behavior. From the phenomenological RCSJ model of a JJ (Resistively and Capacitively Shunted Junction)[17], the total current through the junction $I$ is determined by $I = I_F + I_{Cap} + I_R + I_J$, where $I_F$ is the fluctuation current of the noise channel, $I_{Cap}$ is the current of the capacitive channel, $I_R$ is the additional resistive channel arising from finite temperature quasiparticle excitation, and $I_J$ is the Josephson current. Since $I_F$, $I_{Cap}$ and $I_R$ have negligible influence on the critical current (see SI Section 5 for detailed discussion), we speculate that the critical current through the JJ should be governed by the Josephson current $I_J$, ($I_J = I_c \sin\varphi$, $I_c = V_c/R_N$ and $\varphi$ is phase difference of two superconductors). Since $\Delta I_c$ is suppressed with $\boldsymbol{B}$ and $I_{c+}$ ($I_{c-}$) are symmetric with $\boldsymbol{B}$, and the current is parallel to the symmetry breaking direction while perpendicular to magnetic field direction in our junction, magnetochiral anisotropy can be ruled out as the driving mechanism here. Also to exclude the $NbSe_2$ electrodes as the driving factor, we fabricated JJs of arbitrarily rotated $NbSe_2/NbSe_2$ and $NbSe_2$/few-layer graphene (FLG)/$NbSe_2$ junctions[34,35], and found they both show a field-induced superconducting diode effect with antisymmetric $\Delta I_c$ vs. $B$ characteristics (see SI Section 3). Therefore $\Delta I_c$ must be induced by asymmetric Josephson tunnelling at zero field in the $Nb_3Br_8$ JJs.

Recently, Kitamura *et al.* discussed non-reciprocal Landau-Zener tunneling in $\mathcal{P}$-breaking materials with polarization[36], explaining that the effect in the normal state originates from direction dependent modulation of the electron tunneling probability across a tunnel barrier due to a shift in wave-packet position. Alternatively, Hu *et al.* proposed a JD based on an interface of p and n-doped superconductors[2] that breaks inversion symmetry and forms a polarized Mott insulating region that can be suppressed by voltage in one direction and elongated in the other direction, resulting in diodic behavior.

In the situation of $NbSe_2$/$Nb_3Br_8$/$NbSe_2$ JJs, $Nb_3Br_8$ was recently predicted to be an obstructed atomic insulator with Wannier charge centers symmetrically pinned at the unoccupied inversion centers in between two Br-Nb-Br layers[37,38] (see SI Section 6). The negative charge centers are separated from the positive ones along the $c$ axis of the crystal, and combined with the asymmetric $NbSe_2$/$Nb_3Br_8$ interfaces, an out-of-plane polarization can occur[38]. In analogy to the previously described theoretical works of polarized systems, we hypothesize that a polarization in the $NbSe_2$/$Nb_3Br_8$/$NbSe_2$ Josephson junctions may induce asymmetric Josephson tunneling and lead to the field-free Josephson diode effect; further theoretical and experimental study is necessary to fully elucidate the mechanism.

In summary, we have realized, for the first time, a Josephson diode with magnetic field free, unidirectional superconductivity and Josephson coupling of supercurrent across the barrier. Using a heterostructure of 2D flakes of vdW materials $NbSe_2$ and $Nb_3Br_8$, we demonstrated durable superconducting half-wave rectification with rectification ratios of at least $10^4$, with a switching current density as low as $2.2 \times 10^2$ A/cm². Optimization of this JD architecture through tuning the geometry or type of the barrier can result in more ideal diodic behavior. However, creating further vdW JJ architectures, like demonstrated here, can uncover emergent phenomena when integrating quantum material barriers leading to novel or anomalous Josephson phenomena through the effect of the intrinsic properties of the barrier. Future work combining materials possessing properties such as topological states, ferroelectricity, magnetoelectricity, noncollinear magnetism, obstructed atomic insulator and emergent properties from twisted heterostructures, with superconducting electrodes may allow the realization of novel Josephson phenomena. Thus the opportunities for creating superconducting electronics utilizing quantum materials are vast and just beginning.

## Online content

Any methods, additional references, Nature Research reporting summaries, source data, extended data, Supplementary information, acknowledgements, peer review information; details of author contributions and competing interests; and statements of data and code availability are available at

## References

1 Kou Misaki & Nagaosa, N. Theory of the nonreciprocal Josephson effect. *Phys. Rev. B* **103**, 245302 (2021).

2 Hu, J., Wu, C. & Dai, X. Proposed design of a Josephson diode. *Phys. Rev. Lett.* **99**, 067004 (2007).

3 Chen, C.-Z. et al. Asymmetric Josephson effect in inversion symmetry breaking topological materials. *Phys. Rev. B* **98**, 075430 (2018).

4 Tokura, Y. & Nagaosa, N. Nonreciprocal responses from non-centrosymmetric quantum materials. *Nat. Commun.* **9**, 3740 (2018).

5 Fruchart, M., Hanai, R., Littlewood, P. B. & Vitelli, V. Non-reciprocal phase transitions. *Nature* **592**, 363 (2021).

6 Akamatsu, T. et al. A van der Waals interface that creates in-plane polarization and a spontaneous photovoltaic effect. *Science* **372**, 68 (2021).

7 Sze, S. M. & Lee, M.-K. *Semiconductor Devices: Physics and Technology, 3rd Edition*. 3rd edn, (Wiley, 2012).

8 Ando, F. et al. Observation of superconducting diode effect. *Nature* **584**, 373 (2020).

9 Wakatsuki, R. et al. Nonreciprocal charge transport in noncentrosymmetric superconductors. *Sci. Adv.* **3**, e1602390 (2017).

10 Zhang, E. et al. Nonreciprocal superconducting $NbSe_2$ antenna. *Nat. Commun.* **11**, 5634 (2020).

11 Baumgartner, C. et al. Supercurrent rectification and magnetochiral effects in symmetric Josephson junctions. *Nat. Nanotechnol.* DOI: 10.1038/s41565-021-01009-9 (2021).

12 Yuan, N. F. Q. & Fu, L. Supercurrent diode effect and finite momentum superconductivity. arXiv: 2106.01909v2 (2021).

13 Daido, A., Ikeda, Y. & Yanase, Y. Intrinsic Superconducting Diode Effect. arXiv:2106.03326 (2021).

14 Ideue, T. et al. Bulk rectification effect in a polar semiconductor. *Nat. Phys.* **13**, 578 (2017).

15 Ideue, T., Koshikawa, S., Namiki, H., Sasagawa, T. & Iwasa, Y. Giant nonreciprocal magnetotransport in bulk trigonal superconductor $PbTaSe_2$. *Phys. Rev. Research* **2**, 042046(R)

(2020).

16  Wang, Y. et al. Gigantic magnetochiral anisotropy in the topological semimetal $ZrTe_5$. arXiv:2011.03329 (2021).

17  Likharev, K. K. *Dynamics of Josephson junctions and circuits*. (Gordon and Breach Science Publishers, 1986).

18  Likharev, K. K. Superconducting weak links. *Rev. Mod. Phys.* **51**, 101 (1979).

19  Dubos, P. et al. Josephson critical current in a long mesoscopic S-N-S junction. *Phys. Rev. B* **63** (2001).

20  Golubov, A. A., Kupriyanov, M. Y. & Il'ichev, E. The current-phase relation in Josephson junctions. *Rev. Mod. Phys.* **76**, 411 (2004).

21  Lee, G.-H. et al. Graphene-based Josephson junction microwave bolometer. *Nature* **586**, 42 (2020).

22  Walsh, E. D. et al. Josephson junction infrared single-photon detector. *Science* **372**, 409 (2021).

23  Clarke, J. & Wilhelm, F. K. Superconducting quantum bits. *Nature* **453**, 1031 (2008).

24  Likharev, K. K. & Semenov, V. K. RSFQ logic/memory family: a new Josephson-junction technology for sub-terahertz-clock-frequency digital systems. *IEEE Trans. Appl. Supercond.* **1**, 3 (1991).

25  Devoret, M. H. & Schoelkopf, R. J. Superconducting circuits for quantum information: An outlook. *Science* **339**, 1169 (2013).

26  Jiang, J. et al. Exploration of new ferromagnetic, semiconducting and biocompatible $Nb_3X_8$ (X = Cl, Br or I) monolayers with considerable visible and infrared light absorption. *Nanoscale* **9**, 2992 (2017).

27  Pasco, C. M., El Baggari, I., Bianco, E., Kourkoutis, L. F. & McQueen, T. M. Tunable magnetic transition to a singlet ground state in a 2D van der Waals layered trimerized Kagome magnet. *ACS Nano* **13**, 9457 (2019).

28  Yoon, J. et al. Anomalous thickness-dependent electrical conductivity in van der Waals layered transition metal halide, $Nb_3Cl_8$. *J. Phys.: Condens. Matter* **32**, 304004 (2020).

29  Haraguchi, Y. et al. Magnetic-nonmagnetic phase transition with interlayer charge disproportionation of $Nb_3$ trimers in the cluster compound $Nb_3Cl_8$. *Inorg. Chem.* **56**, 3483 (2017).

30  Sheckelton, J. P., Plumb, K. W., Trump, B. A., Broholm, C. L. & McQueen, T. M. Rearrangement of van der Waals stacking and formation of a singlet state at T = 90 K in a cluster magnet. *Inorg. Chem. Front.* **4**, 481 (2017).

31  Castellanos-Gomez, A. et al. Deterministic transfer of two-dimensional materials by all-dry

viscoelastic stamping. *2D Mater.* **1**, 011002 (2014).

32  Xi, X. et al. Ising pairing in superconducting NbSe$_2$ atomic layers. *Nat. Phys.* **12**, 139 (2015).

33  Kleiner, R., Koelle, D., Ludwig, F. & Clarke, J. Superconducting quantum interference devices: State of the art and applications. *Proc. IEEE* **92**, 1534 (2004).

34  Yabuki, N. et al. Supercurrent in van der Waals Josephson junction. *Nat. Commun.* **7**, 10616 (2016).

35  Kim, M. et al. Strong Proximity Josephson Coupling in Vertically Stacked NbSe$_2$-Graphene-NbSe$_2$ van der Waals Junctions. *Nano Lett.* **17**, 6125 (2017).

36  Kitamura, S., Nagaosa, N. & Morimoto, T. Nonreciprocal Landau–Zener tunneling. *Commun. Phys.* **3**, 63 (2020).

37  Xu, Y. et al. Filling-Enforced Obstructed Atomic Insulators. arXiv:2106.10276 (2021).

38  Xu, Y. et al. Three-Dimensional Real Space Invariants, Obstructed Atomic Insulators and A New Principle for Active Catalytic Sites. arXiv: 2111.02433v1 (2021).

## Methods

### Synthesis of Nb₃Br₈ crystal

Nb$_3$Br$_8$ crystals used in this work were grown through chemical vapor transport. Stoichiometric mixtures of niobium powder (Alfa, 99.99%), and NbBr$_5$ (Strem, 99.9%) with a total mass of 1.5 g were ground together and added to a 14×16 mm diameter fused silica tube in a glovebox and handled using standard air free techniques. 20 mg of NH$_4$Br was included as a transport agent. The tubes were then sealed air free at a length of approximately 30 cm, about 5 cm longer than the first two zones of a three-zone furnace. A three-zone furnace was used with a temperature gradient of 840°C, 785°C and 795°C with all but the last few centimeters of the tube between the first two zones. This discouraged the formation of large intergrown clumps of crystals at the end of the tube. The furnace was held at temperature for 3-5 days before being cooled to room temperature over 7 hours.

### Fabrications of the Josephson junction device.

We fabricated the NbSe$_2$/Nb$_3$Br$_8$/NbSe$_2$ Josephson junction in glove box with inert environment to avoid oxidation and decay. The bottom NbSe$_2$ flake was directly exfoliated from an NbSe$_2$ single crystal (HQ graphene) to a SiO$_2$/Si wafer, then Nb$_3$Br$_8$ thin film and top NbSe$_2$ flake were exfoliated and successively transferred layer by layer by using PDMS assistant dry transfer method[27]. A h-BN (HQ graphene) layer was finally transferred on the top to protect the JJ from degradation in the atmosphere. Then Ti (3 nm)/Au (50 nm) electrodes were deposited on top and bottom NbSe$_2$ flakes for transport measurement.

### Transport measurements

The transport properties of JJ device were measured in a BlueFors dilution refrigerator with a base temperature of 20 mK. A Keithley 6221 AC/DC Current Source Meter was used to inject both DC current and square-wave excitation, and the DC voltage was measured with a Keithley 2182a Nanovoltmeter. The AC measurements were performed by injecting AC current with frequency of $\omega$ (7.919 Hz) using Zurich MFLI, and the second-harmonic signals were obtained from the out-of-phase $2\omega$ component of the AC voltage. Fraunhofer pattern was obtained by measuring the differential resistance (d$V$/d$I$) vs $I$ curve under different magnetic fields using a lock-in amplifier (Zurich MFLI) with a DC bias (Keithley 2636).

### Excluding the Joule heating effect

In Fig. 1b, we measured positive and negative sweep $V$-$I$ curves with the same sweep rates. These two curves lie on top of each other, indicating that the $V$-$I$ curves and associated $I_{c+}$ and $I_{c-}$ are not dependent on the two sweep directions. As shown in Fig. 1c, the $V$-$I$ is asymmetric and $I_{c+}$ and $I_{c-}$ have prominently different values. In general, the Joule heating effect is proportional to the heating time and the $I_{c+}$ (or $I_{c-}$) obtained in positive and negative sweeps undergo different heating times which can cause different local temperatures in sample. If the asymmetric *V-I* was induced by the Joule heating effect, the values of $I_{c+}$ and $|I_{c-}|$ would switch for the positive and negative sweeps; i.e. the effect would be sweep direction dependant. However, the *V-I* curves of positive and negative sweeps are the same, strongly indicating that the different $I_{c+}$ and $I_{c-}$ obtained indeed are intrinsic to the device rather than induced extrinsically by the Joule heating effect.

**Data availability**

The data that support the findings of this study are available from the corresponding author upon request.

**Acknowledgments:**

We thank Justin Song, Gil-Ho Lee and Yongjian Wang for valuable discussions. **Funding:** M.N.A acknowledges that this research was principally supported by the Alexander von Humboldt Foundation Sofia Kovalevskaja Award, the German Federal Ministry of Education and Research's MINERVA ARCHES Award, the Max Planck Society, and Delft University of Technology. Y.-J.Z. acknowledges Shenzhen Science and Technology Project under grant No. JCYJ20180507182246321. S.S.P.P. acknowledges the European Research Council (ERC) under the European Union's Horizon 2020 research and innovation programme (grant agreement no. 670166), Deutsche Forschungsgemeinschaft (DFG, German Research Foundation)—project number 314790414, and Alexander von Humboldt Foundation in the framework of the Alexander von Humboldt Professorship endowed by the Federal Ministry of Education and Research. T.M. acknowledges the David and Lucile Packard Foundation and the Johns Hopkins University Catalyst Award. E.S.T. acknowledges support of NSF DMR 1555340.

**Author contributions:** H.W. and M.N.A. conceived and designed the study. C.P. grew the samples. H.W. and Y.W. fabricated the devices. H.W., Y.W., P.K.S. and U.F. performed the transport measurements. H.W. and Y.W carried out the data analysis. Y.X. provided theoretical support and discussion. T.M. and M.N.A are the Principal Investigators. All authors contributed to the preparation of manuscript.

**Competing interests:** The authors declare that they have no competing interests.

**Additional information**

**Correspondence and requests for materials** should be addressed to H.W. and M.N.A.

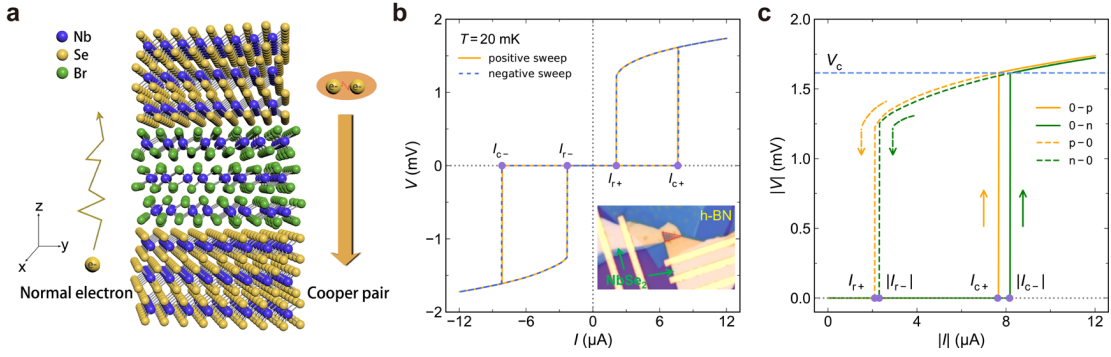

**Fig. 1 | Schematic and superconductivity of the Josephson diode at zero field. a**, schematic of Josephson diode in a vertical Josephson Junction architecture, consisting of an NbSe$_2$/Nb$_3$Br$_8$/NbSe$_2$ sandwich. The junction lacks inversion symmetry along the $z$-direction, and can exhibit normal conduction in one direction but superconductivity in the other (cooper pair conduction). **b**, Voltage vs current curves measured by sweeping positive (orange solid line) and sweeping negative (blue dashed line), as defined in the main text. Inset shows a typical device architecture, the red triangle outlines the Josephson junction geometry created by the overlapping flakes. **c**, Voltage vs current curves with positive sweep; the orange solid and dashed lines correspond with the 0-p and p-0 branches, respectively, while the green solid and dashed lines correspond with the 0-n and n-0 branches, respectively. The purple dots mark the position of $I_{c+}$, $|I_{c-}|$, $I_{r+}$ and $|I_{r-}|$ and the blue dashed line shows the $V_c$.

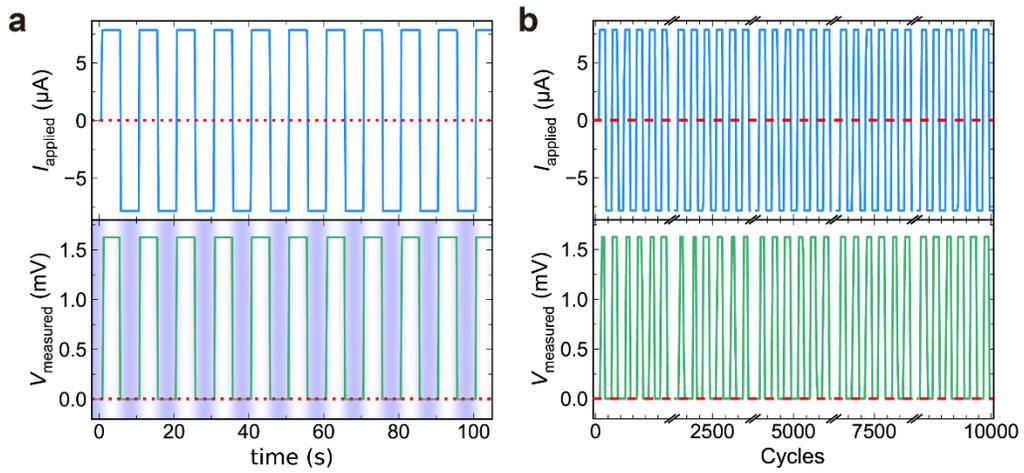

**Fig. 2 | Half-wave rectification and durability test of the Josephson diode at 20 mK and zero-field.**
**a**, Top panel is the applied square-wave excitation with an amplitude of 7.9 μA (in between $I_{c+}$ and $|I_{c-}|$) and frequency of 0.1 Hz. The bottom panel is the coincidentally measured junction voltage. The purple shaded region denotes the superconducting state where the voltage is 0 during negative current bias and the white region denotes the normal state where the voltage is $V_c$ during positive current bias. The red dotted line is the zero line. **b**, The top panel is the applied square-wave excitation with an amplitude of 7.9 μA and frequency of 0.5 Hz over 10,000 cycles. The bottom panel is the coincidentally measured junction voltage over those cycles, showing no degradation in the response. The red dashed lines are zero lines.

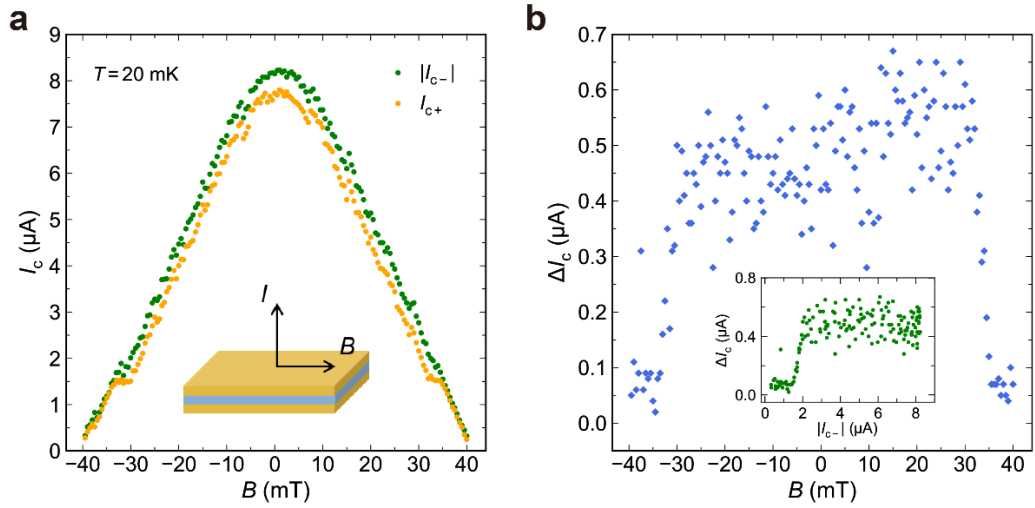

**Fig. 3 | Magnetic field dependence of $I_c$ and $\Delta I_c$. a**, $I_{c+}$ (orange dots) and $|I_{c-}|$ (green dots) obtained from the 0-p and 0-n branches of the positive sweep as a function of applied magnetic field. The magnetic field was swept from positive to negative. The diode effect "turns off" by ±35 mT. Inset is a schematic of the measurement geometry, the orange and blue layers represent NbSe$_2$ and Nb$_3$Br$_8$, respectively. **b**, $\Delta I_c$ as a function of magnetic field. $\Delta I_c$ is roughly field invariant until 35 mT with the diode behavior robust in zero field. Inset shows $\Delta I_c$ as a function of $|I_{c-}|$ from modulation of magnetic field with the diode effect "turning off" below 2.1 µA.

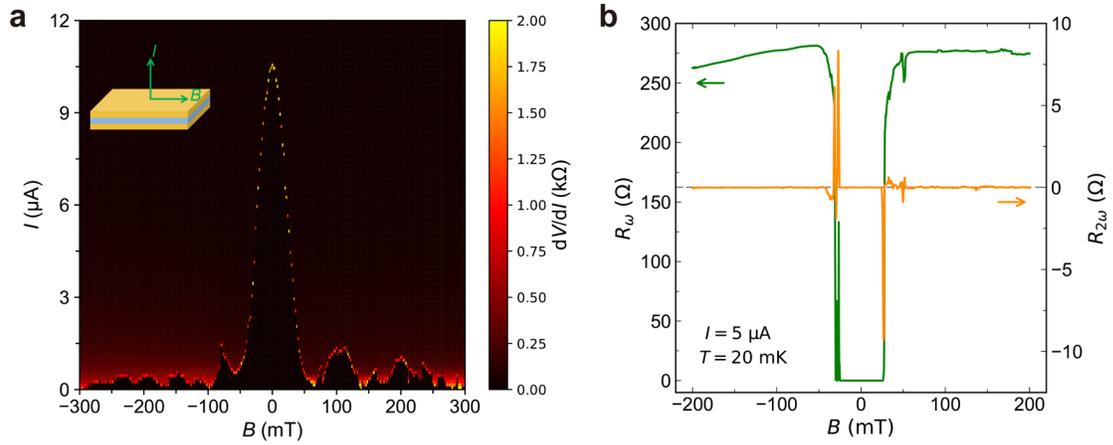

**Fig. 4 | Critical current map and second harmonic resistance of the Josephson diode. a**, Color map of d$V$/d$I$ as a function of applied magnetic field and applied current, showing the single slit Fraunhofer pattern from the Josephson effect of a Josephson junction. **b**, First (green line) and second harmonic (orange line) resistances measured by applying AC current of 5 μA at 20 mK. The antisymmetric peaks in the second harmonic correspond with the critical field, showcasing the nonreciprocal nature of the superconductivity.

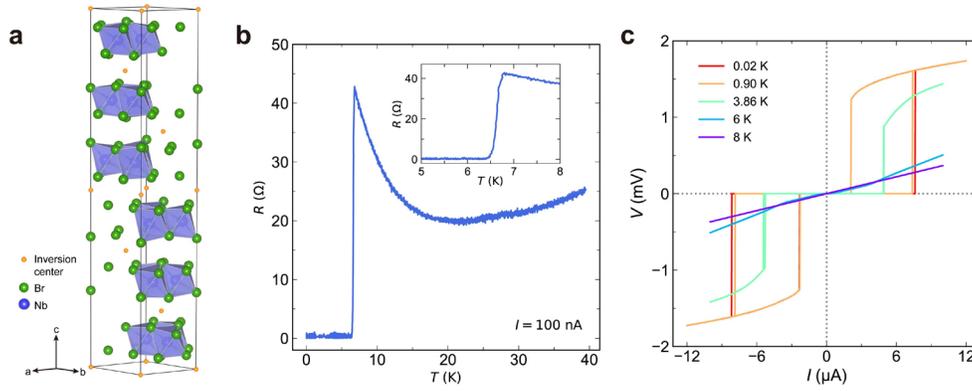

**Extended Data Fig. 1 | Temperature dependent resistance and *V-I* curves of the Josephson diode.**
**a**, Nb$_3$Br$_8$ crystal structure and inversion center locations. **b**, Resistance vs. temperature measured using an AC current of 100 nA. Inset shows the enlarged plot near the superconducting transition of 6.6 K. **c**, *V-I* curves with positive sweep measured at different temperatures showing nonlinear behavior appearing below *T*c.

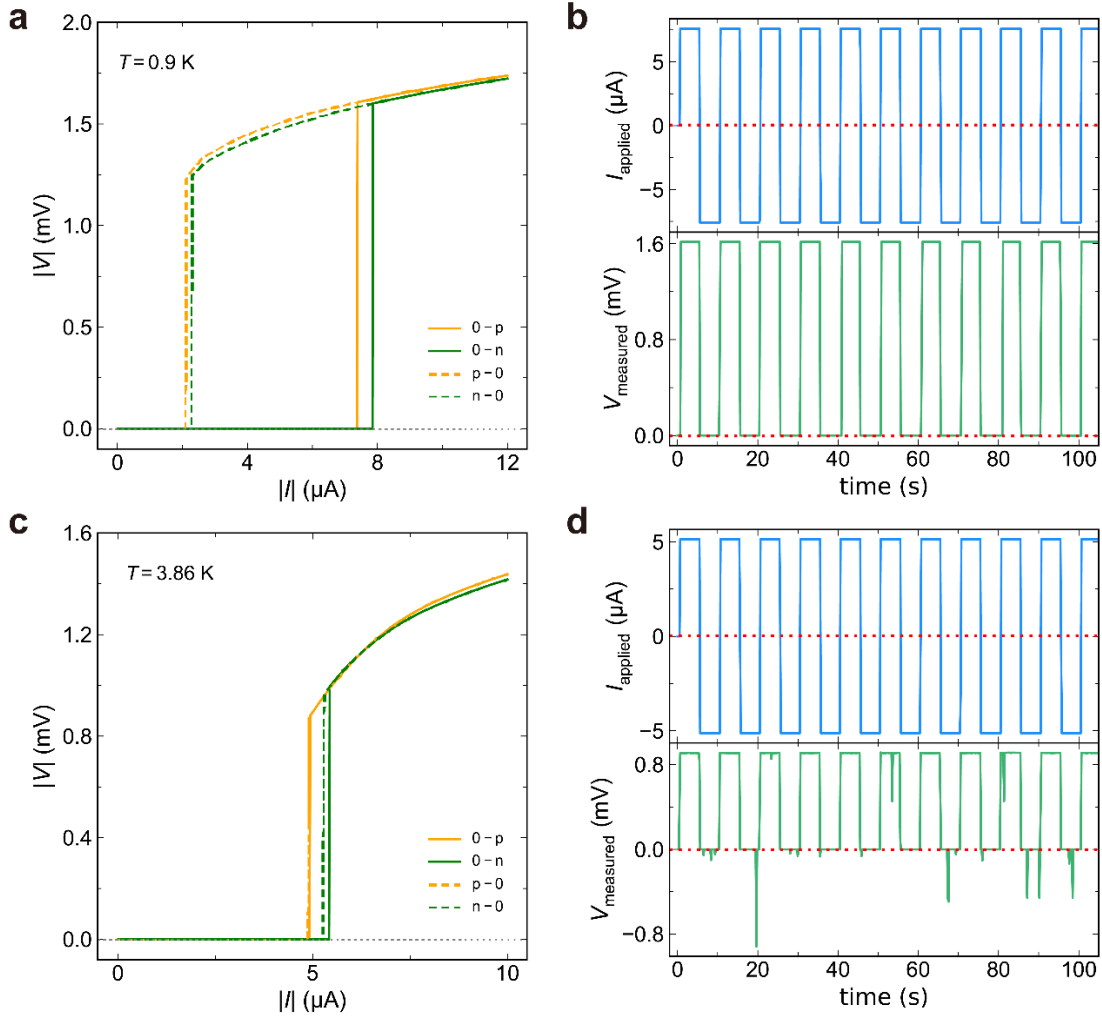

**Extended Data Fig. 2 | *V-I* curves and half-wave rectification at different temperatures. a**, Positive sweep *V-I* curve measured at 0.9 K. Both $\Delta I_c$ and $\Delta I_r$ are visible. **b**, Half-wave rectification measured at 0.9 K with an applied current of 7.6 μA at 0.1 Hz. The red dotted lines are the zero lines, showing that junction is in the superconducting state with negative current and switches to the normal state with positive current. **c**, Positive sweep *V-I* curve measured at 3.86 K. $\Delta I_c$ is still visible, while the hysteresis is almost completely suppressed. **d**, Half-wave rectification measured at 3.86 K with an applied current of 5.14 μA at 0.1 Hz. The red dotted lines are the zero lines. Imperfect rectification is evident with some punch through error, likely due to thermal fluctuation.

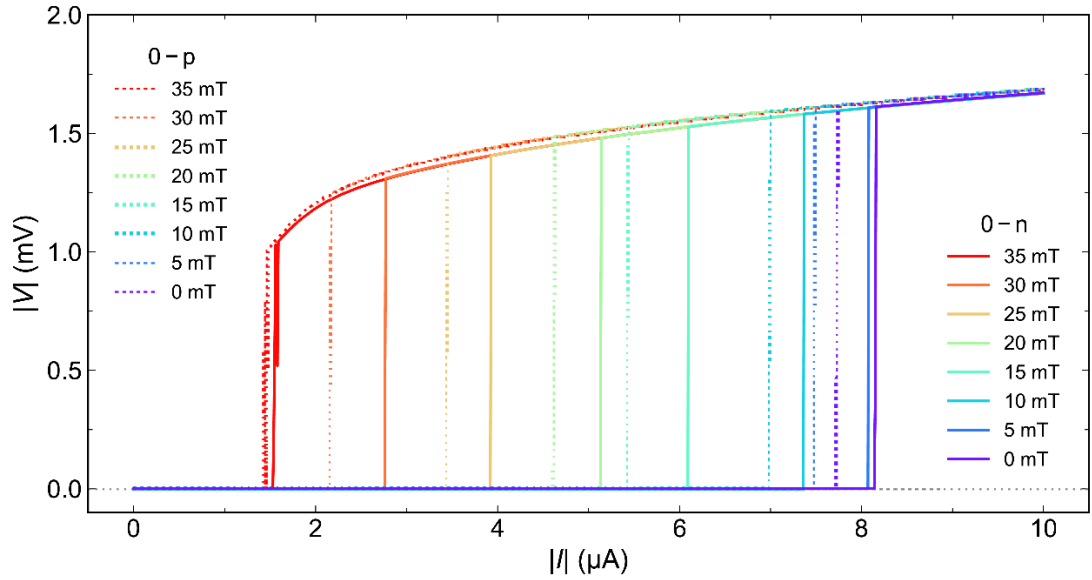

**Extended Data Fig. 3 | *V-I* curves with 0-p and 0-n branches measured at different magnetic fields.** Solid lines are 0-n branches (where $I_{c\text{-}}$ was extracted) and dotted lines are 0-p branches (where $I_{c+}$ was extracted) corresponding to Fig. 3. $\Delta I_c$ almost "turns off" at 35 mT.

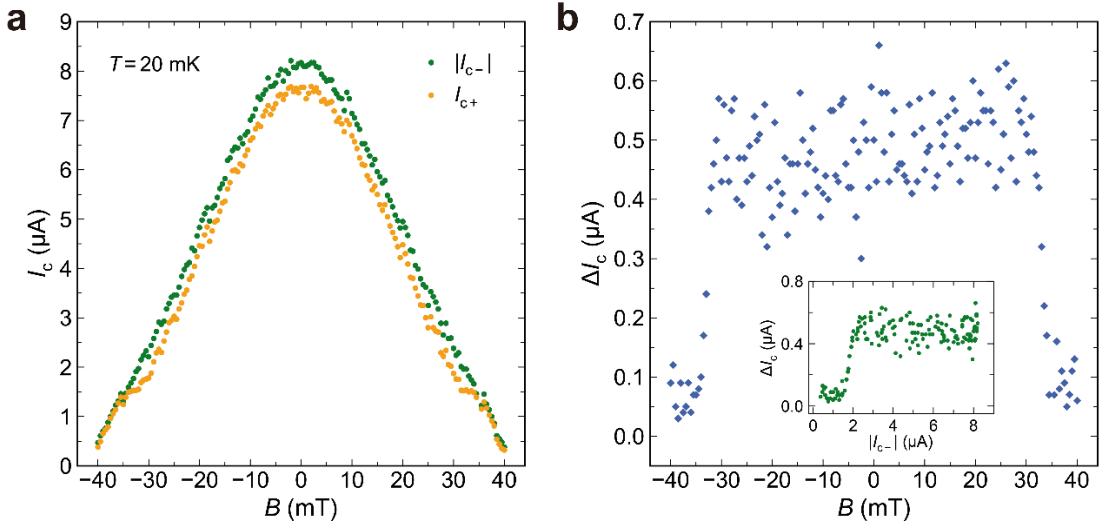

**Extended Data Fig. 4 | Sweep-up magnetic field dependence of $I_c$ and $\Delta I_c$. a,** $I_{c+}$ (orange dots) and $|I_{c-}|$ (green dots) obtained from the 0-p and 0-n branches of the positive sweep as a function of applied magnetic field. The in-plane magnetic field was swept from negative to positive. **b,** $\Delta I_c$ as a function of magnetic field. Inset shows $\Delta I_c$ as a function of $|I_{c-}|$ from modulation of magnetic field with the diode effect "turning off" below 2.1 μA. These sweep-up results are nearly identical with the sweep-down results shown in the main text Fig. 3.

Heng Wu[1,2,4,6*], Yaojia Wang[1,4,6], Yuanfeng Xu[1,5], Pranava K. Sivakumar[1], Chris Pasco[3], Ulderico Filippozzi[4], Stuart S. P. Parkin[1], Yu-Jia Zeng[2], Tyrel McQueen[3], Mazhar N. Ali[1,4*]

[1]Max Planck Institute of Microstructure Physics, Halle, Saxony-Anhalt, 06108, Germany

[2]College of Physics and Optoelectronic Engineering, Shenzhen University, Shenzhen 518060, China

[3]Johns Hopkins University, Baltimore, Maryland 21218, USA

[4]Kavli Institute of Nanoscience, Delft University of Technology, Delft, the Netherlands

[5]Department of Physics, Princeton University, Princeton, New Jersey 08544, USA

[6]These authors contributed equally: Heng Wu, Yaojia Wang

*e-mail: wuhenggcc@gmail.com, maz@berkeley.edu

# Contents

**Section 1. Magnetoresistance characterization of Nb₃Br₈ thin flake.**

The Nb₃Br₈ single crystal has been studied to show a phase transition from a high temperature ferromagnetic state to a magnetic singlet ground state at low temperature ($\beta$ phase)[1]. Here, we use transport measurements to investigate the magnetic property of Nb₃Br₈ thin flake. First, we measured sweep-up and sweep-down magnetoresistances of NbSe₂/Nb₃Br₈/NbSe₂ JJ (Device #1, device in main text) with temperature above $T_c$ of NbSe₂ superconducting electrodes. Figure S1a and S1b show the magnetoresistances of Device #1 at 10 K with in-plane (IP) and out-of-plane (OOP) magnetic fields up to 7 T, respectively. The overlapping sweep-up and sweep-down magnetoresistances curves reveal that there are no hysteresis or reversion between these curves, indicating no ferromagnetic signal in the Nb₃Br₈ thin flakes[2,3]. Moreover, we also replaced NbSe₂ superconducting electrodes with few-layer graphene (FLG) to fabricated FLG/Nb₃Br₈/FLG devices (Device #6) and measured their magnetoresistances with OOP magnetic field. Figure S1c shows the results of one of the FLG/Nb₃Br₈/FLG devices (Device #6) measured at 2K. Similar to the results in Device #1, the magnetoresistances are overlapping well, and no hysteresis or reversion signals can be observed in the curves, which is also consistent with the observation of the magnetic singlet ground state in ref. 1. It's worth noting that the oscillations in FLG/Nb₃Br₈/FLG devices are induced by the quantum oscillation of FLG.

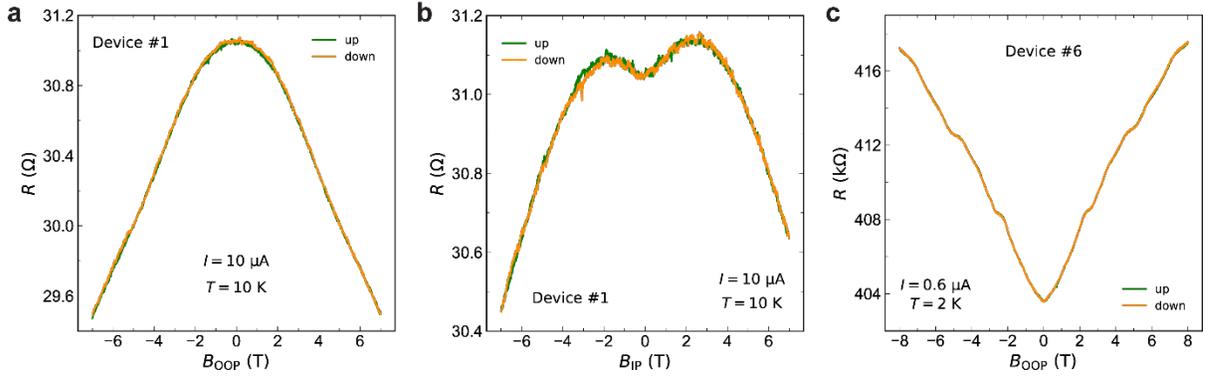

**Fig. S1** | **a**, Sweep-up and sweep-down magnetoresistances of NbSe₂/Nb₃Br₈/NbSe₂ (Device #1) with out-of-plane magnetic field at 10 K (above $T_c$ of NbSe₂). **b**, Sweep-up and sweep-down magnetoresistances of Device #1 with in-plane magnetic field at 10 K. **c**, Sweep-up and sweep-down magnetoresistance of FLG/Nb₃Br₈/FLG (Device #6) with out-of-plane magnetic field at 2 K.

**Section 2. Field-free Josephson diode effect in a NbSe₂/4-layer Nb₃Br₈/NbSe₂ Josephson junction (Device #3)**

NbSe₂/Nb₃Br₈/NbSe₂ Josephson junction with a 4-layer Nb₃Br₈ as the tunnel barrier (Device #3) was fabricated with the same technique as Device #1 (device in main text). We measured $V$-$I$ curves at different magnetic fields at 20 mK and extracted $I_{c+}$ and $|I_{c-}|$ from 0-p and 0-n branches, respectively, and plotted as a function of magnetic field as shown in Fig. S2a. The $I_{c+}$ and $|I_{c-}|$ exhibit similar tendency and behavior with those in Device #1. Figure S2b shows the field dependent calculated $\Delta I_c$. The non-zero $\Delta I_c$ can be clearly observed at zero field and low field region, and it decreases to almost 0 at relative higher fields which is similar with the phenomenon in Device #1. Therefore, the JJ with 4-layer Nb$_3$Br$_8$ also has field-free Josephson diode effect. To confirm the symmetry broken, we also performed the second harmonic resistances ($R_{2\omega}$) measurements at 20 mK with applied currents of 4 μA, as shown in Fig. S2c. The two antisymmetric $R_{2\omega}$ peaks corresponding with the superconducting transition of $R_\omega$ reveal the $\mathcal{P}$-breaking nature of the JJ[4,5]. Since the 4-layer Nb$_3$Br$_8$ preserves inversion symmetry as shown in Extended Data Figure 1a, the $\mathcal{P}$-breaking of Device #3 origins from the different interfaces of top and bottom NbSe$_2$/Nb$_3$Br$_8$.

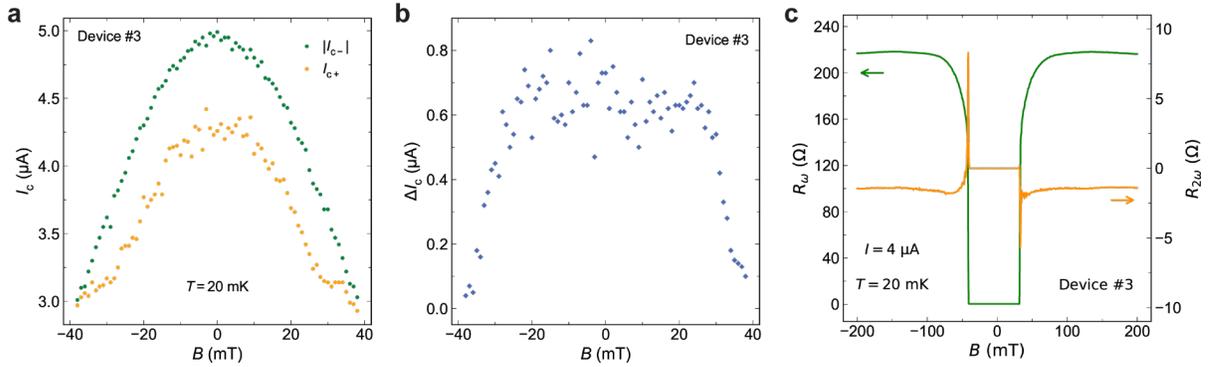

**Fig. S2 | a,** $I_{c+}$ and $|I_{c-}|$ as a function of magnetic field at 20 mK in NbSe$_2$/4-layer Nb$_3$Br$_8$/NbSe$_2$ junction (Device #3). **b,** $\Delta I_c$ as a function of magnetic field of Device #3. **c,** First harmonic ($R_\omega$) and second harmonic resistances ($R_{2\omega}$) as a function of magnetic field measured at 20 mK with an applied current of 4 μA of Device #3.

## Section 3. Magnetic field-induced superconducting diode effects in NbSe$_2$/NbSe$_2$ (Device #4 and #7) and NbSe$_2$/FLG/NbSe$_2$ (Device #5) junctions

We also fabricated other junctions as control experiments to narrow the suspects of the field-free Josephson diode effect mechanism. First, we consider the influence of the superconducting electrode NbSe$_2$, which is a type II and multi-gapped superconductor. Previous study reported the nonreciprocal transport in atomically thin NbSe$_2$ films that break inversion symmetry, which disappear in bulk NbSe$_2$ crystal[6]. In all our devices, we use thick NbSe$_2$ flakes (>20 nm) as the superconducting electrodes that have similar property with bulk NbSe$_2$. We fabricated NbSe$_2$/NbSe$_2$ junction (Device #4) to explore its property. Figure S3a shows the $V$-$I$ curve measured with the positive sweep (a sequence of 0-p, p-0, 0-n and n-0 branches) at 2 K and 0 T. As indicated in Fig. S3a, there are 3 transition steps in the $V$-$I$ curve, which correspond with the critical currents of junction, bottom and top NbSe$_2$ electrodes, respectively. There's no obvious difference between the positive critical current ($I_{c+}$) and negative one ($I_{c-}$) of the junction. The field dependence of critical currents is further measured, we extracted the $I_{c+}$ and $|I_{c-}|$ of the junction from each $V$-$I$ curve measured at different magnetic fields and plotted as a function of magnetic field as shown in Fig. S3b. The $I_{c+}$ and $|I_{c-}|$ seem to have a mirror symmetry with each other. $\Delta I_c$ was also calculated and shown in Fig. S3c. The antisymmetric $\Delta I_c$ curve indicates that NbSe$_2$/NbSe$_2$ junction can only exhibit field-induced superconducting diode effect, rather than field-free Josephson diode effect observed in NbSe$_2$/Nb$_3$Br$_8$/NbSe$_2$ junctions. To further confirm the influence of barrier materials, we also fabricated NbSe$_2$/FLG/NbSe$_2$ junction (Device #5) as a control experiment. The $I_{c+}$ and $|I_{c-}|$ was extracted as before and shown in the inset of Fig. S3d. $\Delta I_c$ was also calculated and plotted as a function of magnetic field in Fig. S3d. Similar with the results of NbSe$_2$/NbSe$_2$ junction device (Device #4), NbSe$_2$/FLG/NbSe$_2$ junction also exhibits the field-induced superconducting diode effect.

Previous studies have shown that a self-field effect in tunnel junctions with a large critical current density can result in a skewed Fraunhofer pattern, which may contribute to the field-induced superconducting diode effect in these junctions. We also fabricated a NbSe$_2$/NbSe$_2$ control device (Device #7) with low critical current ($I_c \sim 120$ μA) and a correspondingly lower critical current density ($J_c \sim 18.5$ μA/μm$^2$) than Device #4, $J_c \sim 66$ μA/μm$^2$. Figure S4 shows the $I_c$ and corresponding $\Delta I_c$ as a function of applied magnetic field. The $I_{c+}$ and $|I_{c-}|$ vs $B$ have a mirror symmetry with each other, and the obvious antisymmetric feature of $\Delta I_c$ vs $B$ curve also confirms the field-induced superconducting diode effect in Device #7, similar to Device #4.

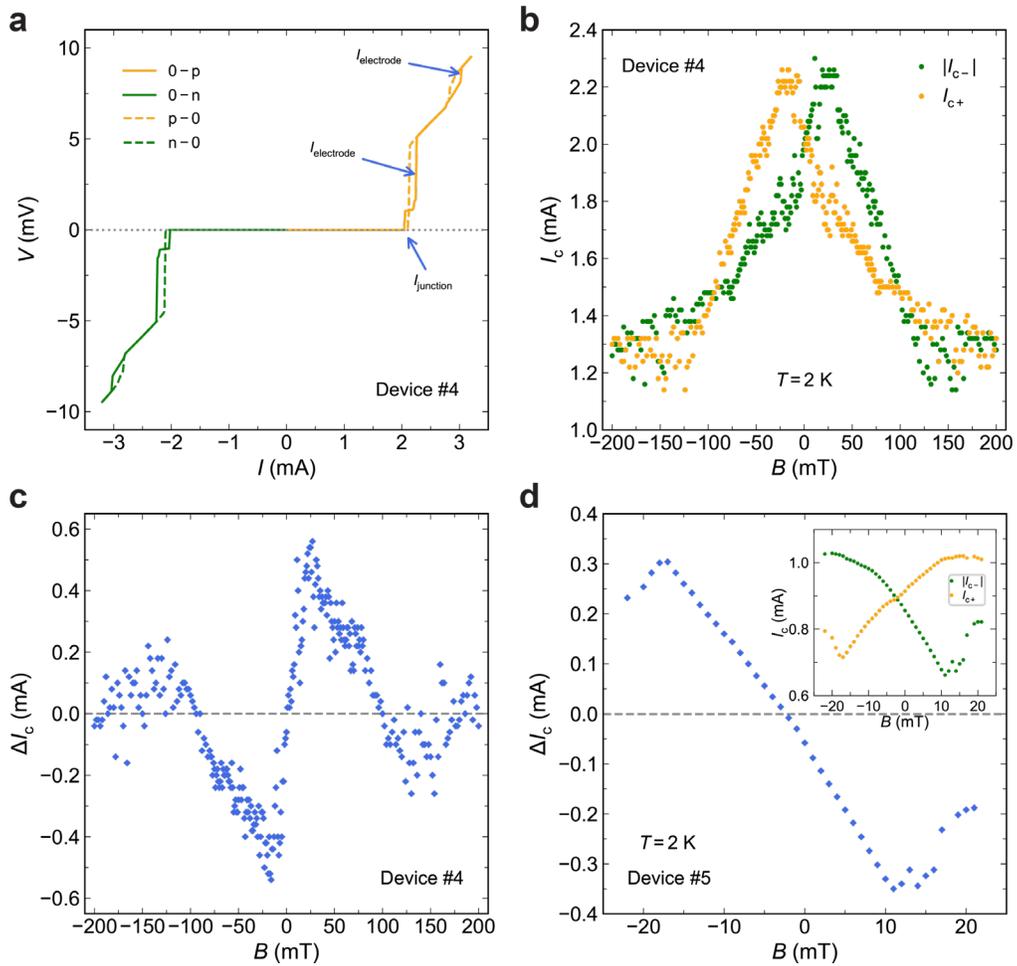

**Fig. S3** | **a**, $V$-$I$ curve of NbSe$_2$/NbSe$_2$ junction (Device #4) measured at 2 K and 0 T. **b**, $I_{c+}$ and $|I_{c-}|$ as a function of magnetic field of Device #4 at 2 K. **c**, $\Delta I_c$ as a function of magnetic field of Device 4. **d**, $\Delta I_c$ as a function of magnetic field of NbSe$_2$/FLG/NbSe$_2$ junction (Device #5) measured at 2 K, inset shows the corresponding $I_{c+}$ and $|I_{c-}|$.

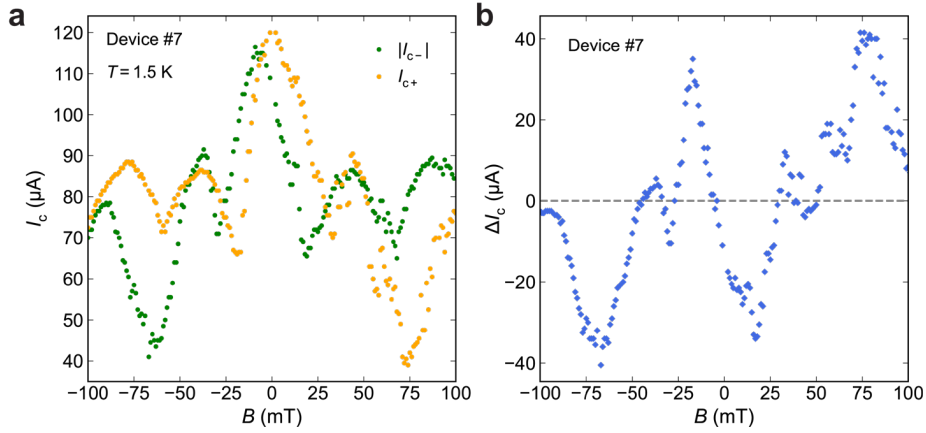

**Fig. S4** | **a**, $I_{c+}$ and $|I_{c-}|$ as a function of applied magnetic field of NbSe$_2$/NbSe$_2$ junction (Device #7). **b**, $\Delta I_c$ as a function of magnetic field of Device #7.

In addition, recent studies on planar Nb/WTe$_2$/Nb Josephson junctions[7] and the InAs 2DEG system[8], in which the self-field effect doesn't exist, also reported a mirrored $I_{c+}$ ($|I_{c-}|$) vs. $B$ curves, similar with the results of these NbSe$_2$/NbSe$_2$ and NbSe$_2$/FLG/NbSe$_2$ junctions. So far, the field-induced superconducting diode effect has been proposed to arise from effects such as magnetochiral anisotropy[5], finite momentum superconductivity[9], and nonreciprocal Landau critical momentum[10]. More investigation is needed to understand the origin of field-induced diode effect in NbSe$_2$/NbSe$_2$ and NbSe$_2$/FLG/NbSe$_2$ junctions.

In summary, both NbSe$_2$/NbSe$_2$ and NbSe$_2$/FLG/NbSe$_2$ junctions exhibit field-induced superconducting diode effect, similar to the behavior of the [Nb/V/Ta]$_n$ superlattice[11]. These observations in NbSe$_2$/NbSe$_2$ and NbSe$_2$/FLG/NbSe$_2$ junctions indicate that the field-free Josephson diode effect does not origin from the superconducting property of NbSe$_2$. And even though the interfaces of these junctions also break the inversion symmetry, the field-free Josephson diode effect dose not occur. These results, along with the observations in NbSe$_2$/Nb$_3$Br$_8$/NbSe$_2$ devices (Device #1 and Device #3), indicate that the field-free Josephson diode effect originates from asymmetric Josephson tunnelling induced by the Nb$_3$Br$_8$ barrier and the associated NbSe$_2$/Nb$_3$Br$_8$ interfaces in the junction.

**Section 4. Basic parameters of the Josephson junction (Device #1) and discussion of Fraunhofer**

**patterns (Device #1 and #2)**

The large hysteresis between 0-p (0-n) and p-0 (n-0) branches on $V$-$I$ curves (Fig. 1b) indicates the JJ (Device #1) lies in the underdamped region. Based on the critical current ($I_c$) and return current ($I_r$) of $V$-$I$ curve measured at 20 mK, the Stewart-McCumber parameter ($\beta_c \approx \left(\frac{4I_c}{\pi I_r}\right)^2$) NbSe$_2$/Nb$_3$Br$_8$/NbSe$_2$ Josephson junction is 21, which is much larger than 1, further confirming the JJ is in underdamped region[12,13]. Based on equation $\beta_c = \frac{2e}{\hbar} I_c R_N^2 C$ (where $R_N$ is normal state resistance of junction)[12,13], the junction capacitance calculated is $2.0\times10^{-14}$ F, corresponding with specific capacitance (C/A, A $\approx$ 3.68 $\mu$m$^2$) of 0.54 $\mu$F/cm$^2$.

The Fraunhofer pattern in Fig. 4a is measured sweeping the magnetic field from negative to positive (sweep-up). Figure S5a shows the Fraunhofer pattern measured with sweep-down magnetic field, which is similar with the sweep-up one. Both Fraunhofer patterns are not so periodic at higher field, which is a common case as seen in many other two dimensional tunnel Josephson junctions, including the NbSe$_2$/NbSe$_2$ junctions[14] as well as NbSe$_2$/graphene/NbSe$_2$ junctions[15]. In these vertical junctions with very thin barriers, the Fraunhofer pattern can be influenced by many effects, such as magnetic vortices in the superconducting electrodes and the self-field effect[15], which can make it difficult to get an ideal, highly periodic Fraunhofer pattern. In addition, from the theory, the ideal Fraunhofer pattern requires a uniform current profile (rectangular junction area shape and magnetic field parallel with either side of the rectangle). Therefore, the junction shape and magnetic field direction can also lead to a non-ideal Fraunhofer pattern, even in a high quality junction with high quality interfaces[16].

Here, we also characterized more junction parameters by analyzing the Fraunhofer pattern in Fig. S5a at low magnetic field. In a JJ, the London penetration depth ($\lambda$) of NbSe$_2$ can be obtained from $\phi_0 = \Delta B W(d + 2\lambda)$, where $\phi_0$ is the magnetic flux quantum, $\Delta B$ is the period of the oscillation in the Fraunhofer, $W$ is the width of the junction perpendicular to the field direction, and $d$ is the barrier thickness. From the Fraunhofer pattern, the oscillation period is ~59 mT and with $W \approx 3.5$ $\mu$m, and $d \approx$

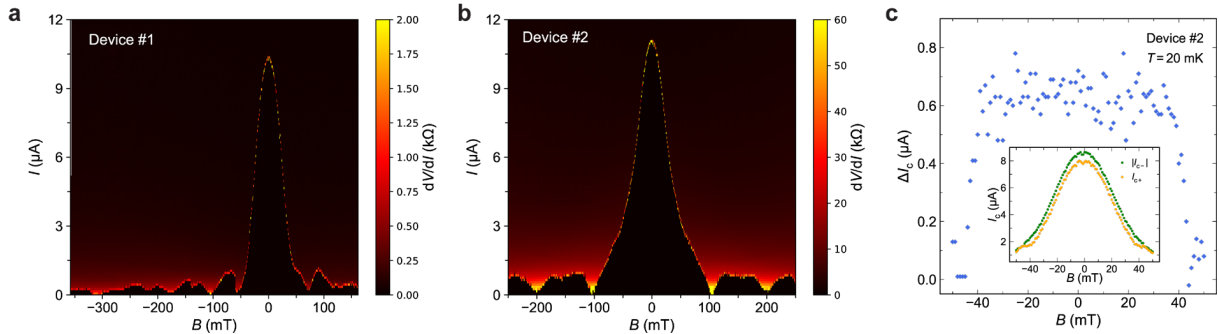

**Fig. S5** | **a**, Fraunhofer pattern of Device #1 with sweep-down magnetic field. **b**, Fraunhofer pattern of Device #2. **c**, $\Delta I_c$ as a function of applied magnetic field of Device #2 measured at 2 K, inset shows the corresponding $I_{c+}$ and $|I_{c-}|$.

2.3 nm, $\lambda$ is calculated to be 3.85nm.

In addition, to further demonstrate the NbSe$_2$/Nb$_3$Br$_8$/NbSe$_2$ heterostructure is indeed a Josephson junction, we fabricated another NbSe$_2$/Nb$_3$Br$_8$/NbSe$_2$ heterostructure device (Device #2), and obtained the expected high-quality Fraunhofer pattern, as shown in Fig. S5b. Moreover, the field-free Josephson diode effect is also observed in this junction, as shown in Fig. S5c and its inset. The $|I_c|$ and $|I_{c-}|$ vs. $B$ curves shown in the inset of Fig. S5c exhibits similar tendency and behavior as with Device #1, and the extracted $\Delta I_c$ exhibit non-zero values at zero field and in the low field region, and decreases to ~0 at higher fields similar to Device #1.

## Section 5. Discussion on RCSJ mode of Josephson diode

Here, we discuss the field-free Josephson diode behavior based on the phenomenological RCSJ model of a JJ (Resistively and Capacitively Shunted Junction)[12], the total current through the junction $I$ is determined by $I = I_F + I_{Cap} + I_R + I_J$, where $I_F$ is the fluctuation current of the noise channel, $I_{Cap}$ is the current of the capacitive channel, $I_R$ is the additional resistive channel arising from finite temperature quasiparticle excitation, and $I_J$ is the Josephson current. In DC measurement, the thermal noise current ($I_T$) dominates the fluctuation current, which can be calculated according to equation $I_T = (2e/\hbar)k_BT$ (ref. [12]), where $\hbar$ is reduced Planck constant, $k_B$ is Boltzmann constant, $e$ is elementary charge, and $T$ is the temperature. At 20 mK, $I_T$ is about 0.8 nA, far less than $\Delta I_c$, and so will be disregarded as a cause of the $V$-$I$ asymmetry. In principle, the other three components could have asymmetric $V$-$I$ responses if inversion/time reversal symmetry is broken. Below we examine each of those possibilities in the context of our JJ.

Regarding the capacitive channel, as mentioned earlier, Misaki et al.[17] proposed that in an inversion symmetry broken JJ, an asymmetric charging energy of capacitance can lead to a $\Delta I_r$, without an applied magnetic field, which was observed in our JJ device (Fig. 1c). However, this model did not include an expectation of a $\Delta I_c$, which is one of the main requirements (the other being sharp transitions) for realizing half-wave rectification. Since $I_r$ corresponds with the return to the superconducting state, applying a square-wave current excitation with an amplitude of $I_{r+} + \Delta I_r/2$, for example, would not result in rectification as both ends of the wave lie within the superconducting regime.

Considering the additional resistive channel ($I_R = V/R_N$) from quasiparticle excitations, its contribution to the diode effect would stem from asymmetry in $R_N$ and be significant if the resistance in the superconducting state was not 0. However, since the voltage in the superconducting state is 0 in our device, the contribution of $I_R$ to the critical current can be neglected. Therefore, the critical current through the JJ is governed by the Josephson current $I_J$, ($I_J = I_c\sin\varphi$, $I_c = V_c/R_N$ and $\varphi$ is phase difference of two superconductors). The difference of positive and negative critical currents should origin from the asymmetric Josephson tunnelling in our JJ.

**Section 6. The obstructed atomic insulator phase of $Nb_3Br_8$**

The crystal structure of $Nb_3Br_8$ has the symmetries of the space group 166 ($R\bar{3}m$), as shown in Extended Data Fig. 1a, where Nb and Br atoms occupy the Wyckoff positions 6c and 18h, respectively. The electronic band structure and the band representation analysis of $Nb_3Br_8$ have been studied[18,19], and the symmetry properties of its valence bands can be characterized by the multiplicities of irreducible representations at all the high-symmetry momenta, which is referred to as the symmetry-data-vector,

$$B = \left(m(\bar{\Gamma}_4\bar{\Gamma}_5), m(\bar{\Gamma}_6\bar{\Gamma}_7), m(\bar{\Gamma}_8), m(\bar{\Gamma}_9), m(\bar{F}_3\bar{F}_4), m(\bar{F}_5\bar{F}_6), ..., m(\bar{T}_9)\right)^T \qquad \text{Eq.(1)}$$
$$= (15,15,33,32,48,47,47,48,15,15,32,33)^T$$

where $m(\rho)$ represents the multiplicity of the irreducible representations $\rho$ formed by the occupied Bloch bands at the corresponding high symmetry momentum. From topological quantum chemistry[20], the band representation, which are characterized by the above symmetry-data-vector, can be expressed as a linear combination of several elementary band representations (EBRs, which are the band representations induced from atomic orbitals) with non-negative-integer coefficients. Hence, $Nb_3Br_8$ is a topologically trivial insulator.

A topologically trivial insulator can be further classified as an atomic insulator or an obstructed atomic insulator (OAI). If the band representation of a topological insulator cannot be induced only from the atomic orbitals at atom occupying sites, there must be some atomic orbitals (or Wannier functions) centered at the empty sites (i.e. the Wyckoff positions that are not occupied by any atoms) and this topologically trivial insulator is referred to as OAI. The OAIs can be indicated by the non-zero integer real space invariants at the empty sites and all the OAIs on the Topological materials database (https://www.topologicalquantumchemistry.org)[20] have been exhausted by performing a high-throughput calculations[19], where $Nb_3Br_8$ is identified as an OAI material indicated by a non-zero real space invariants at the empty Wyckoff position *3b*, whose definition is

$$\delta(b) = \frac{1}{2}[-m(\bar{\Gamma}_6\bar{\Gamma}_7) + m(\bar{\Gamma}_9) + m(\bar{F}_5\bar{F}_6) - m(\bar{L}_5\bar{L}_6) + m(\bar{T}_6\bar{T}_7) - m(\bar{T}_9)]. \quad \text{Eq.(2)}$$

By substituting the symmetry-data-vector in Eq.(1) in to Eq. (2), we have $\delta(b) = -1$, which means there must have a Wannier function centered at *3b*. As the positions of *3b* is not occupied by any atom, $\delta(b) = -1$ indicates that $Nb_3Br_8$ is an OAI and the Wyckoff position *3b* is referred to as the obstructed Wannier charge center (OWCC).

To demonstrate the OAI phase of $Nb_3Br_8$, we have also calculated the electron density distribution in the lattice of $Nb_3Br_8$. As the results shown in Fig. S6, the electron charge density is symmetrically centered at the Wyckoff position *3b* (the inversion center between two $Nb_3Br_8$ layers), indicating the special separation of positive and negative charge centers along the *c* axis of the crystal. In the heterostructures of $NbSe_2/Nb_3Br_8/NbSe_2$, the asymmetric top and bottom interfaces break inversion symmetry of the whole junction along *c* axis. The distribution of charge centers between layers is easy to be influenced by this symmetry breaking, which could result in the charge polarization in the whole junction along *z* direction and hence exhibit an out-of-plane polarization. As discussed in the main text, in analogy to previous theoretical investigations of the polarized system[21,22], we hypothesize that a polarization in the $NbSe_2/Nb_3Br_8/NbSe_2$ Josephson junctions may induce asymmetric Josephson tunnelling and lead to the field-free Josephson diode effect; further theoretical and experimental study is necessary to fully elucidate the mechanism.

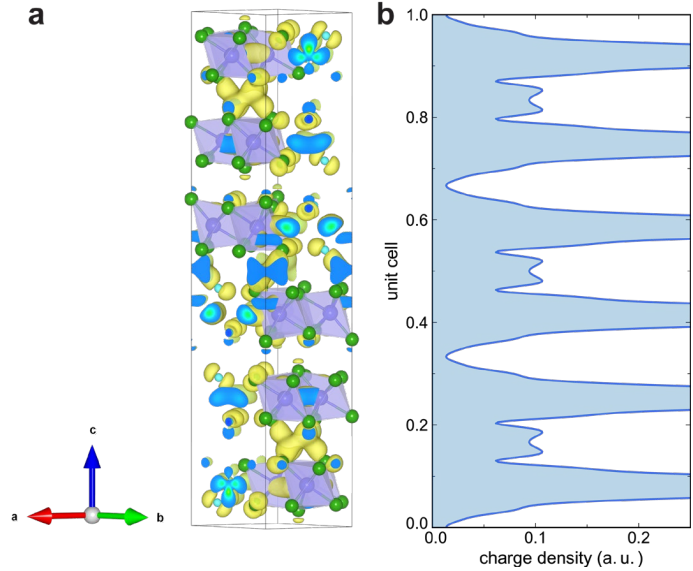

**Fig. S6** | **a**, charge density distribution in a $Nb_3Br_8$ unit cell, the yellow/blue lobes indicate the charge density. **b**, charge density distribution as a function of unit cell location. Note the charge density does not drop to near 0 in every other van der Waals gap.

### References

1    Pasco, C. M., El Baggari, I., Bianco, E., Kourkoutis, L. F. & McQueen, T. M. Tunable magnetic transition to a singlet ground state in a 2D van der Waals layered trimerized Kagome magnet. *ACS Nano* **13**, 9457 (2019).

2    Song, T. et al. Giant tunneling magnetoresistance in spin-filter van der Waals heterostructures. *Science* **360**, 1214 (2018).

3    Wang, Z. et al. Very large tunneling magnetoresistance in layered magnetic semiconductor $CrI_3$. *Nat. Commun.* **9**, 2516 (2018).

4    Ideue, T., Koshikawa, S., Namiki, H., Sasagawa, T. & Iwasa, Y. Giant nonreciprocal magnetotransport in bulk trigonal superconductor $PbTaSe_2$. *Phys. Rev. Research* **2**, 042046(R) (2020).

5    Wakatsuki, R. et al. Nonreciprocal charge transport in noncentrosymmetric superconductors.

*Sci. Adv.* **3**, e1602390 (2017).

6   Zhang, E. et al. Nonreciprocal superconducting NbSe$_2$ antenna. *Nat. Commun.* **11**, 5634 (2020).

7   Kononov, A. et al. One-dimensional edge transport in few-layer WTe$_2$. *Nano Lett.* **20**, 4228 (2020).

8   Baumgartner, C. et al. Supercurrent rectification and magnetochiral effects in symmetric Josephson junctions. *Nat. Nanotechnol.* DOI: 10.1038/s41565-021-01009-9 (2021).

9   Yuan, N. F. Q. & Fu, L. Supercurrent diode effect and finite momentum superconductivity. arXiv: 2106.01909v2 (2021).

10  Daido, A., Ikeda, Y. & Yanase, Y. Intrinsic Superconducting Diode Effect. arXiv:2106.03326 (2021).

11  Ando, F. et al. Observation of superconducting diode effect. *Nature* **584**, 373 (2020).

12  Likharev, K. K. *Dynamics of Josephson junctions and circuits*. (Gordon and Breach Science Publishers, 1986).

13  Likharev, K. K. Superconducting weak links. *Rev. Mod. Phys.* **51**, 101 (1979).

14  Yabuki, N. et al. Supercurrent in van der Waals Josephson junction. *Nat. Commun.* **7**, 10616 (2016).

15  Kim, M. et al. Strong Proximity Josephson Coupling in Vertically Stacked NbSe$_2$-Graphene-NbSe$_2$ van der Waals Junctions. *Nano Lett.* **17**, 6125 (2017).

16  Watanabe, N. et al. The shape dependency of two-dimensional magnetic field dependence of a Josephson junction. *J. Appl. Phys.* **103**, 07C707 (2008).

17  Kou Misaki & Nagaosa, N. Theory of the nonreciprocal Josephson effect. *Phys. Rev. B* **103**, 245302 (2021).

18  Vergniory, M. G. et al. A complete catalogue of high-quality topological materials. *Nature* **566**, 480 (2019).

19  Xu, Y. et al. Three-Dimensional Real Space Invariants, Obstructed Atomic Insulators and A New Principle for Active Catalytic Sites. arXiv: 2111.02433v1 (2021).

20  Bradlyn, B. et al. Topological quantum chemistry. *Nature* **547**, 298 (2017).

21  Kitamura, S., Nagaosa, N. & Morimoto, T. Nonreciprocal Landau–Zener tunneling. *Commun. Phys.* **3**, 63 (2020).

22  Hu, J., Wu, C. & Dai, X. Proposed design of a Josephson diode. *Phys. Rev. Lett.* **99**, 067004 (2007).